\documentstyle[fl eqn,twoside]{article}
\def\be{\begin{equation}}
\def\ee{\end{equation}}
\def\bi{\bibitem}

\begin{document}
\title{Noether And Some Other Dynamical Symmetries In Kantowski-Sachs Model}
\author{ABHIK KUMAR SANYAL}

\maketitle

\noindent

\begin{center}
Dept. of Physics, Jangipur College, Murshidabad,
\noindent
India - 742213\\
\noindent
and\\
\noindent
Relativity and Cosmology Research Centre\\
\noindent
Dept. of Physics, Jadavpur University\\
\noindent
Calcutta - 700032, India\\
\noindent
e-mail : aks@juphys.ernet.in\\
\end{center}

\noindent
\begin{center}
\bf{\Large{Abstract}}
\end{center}                    

The forms of coupling of the scalar field with gravity, appearing in the 
induced theory of gravity, and the potential are found in the 
Kantowski-Sachs model under the assumption that the Lagrangian admits 
Noether symmetry. The form thus obtained makes the Lagrangian 
degenerate. The constrained dynamics thus evolved due to such degeneracy 
has been analysed and a solution has also been presented which is 
inflationary in behaviour.It has further been shown that there exists 
other technique to explore the dynamical symmetries of the Lagrangian and 
that is simply by inspecting the field equations.Through this method 
Noether along with some other dynamical symmetries are found 
which do not make the Lagrangian degenerate.


\section{\bf{Introduction}}

The theory of induced gravity has been found to be a strong candidate in 
several unified theories \cite {m:c}, in which the Einstein-Hilbert action 
appears as an effective action induced by the quantum properties of the 
vacuum state of matter field in the weak energy limit. The beauty of the 
theory lies in the fact that it identifies the inflaton with the scalar 
field inducing the Newtonian gravitational constant$(G_{N})$ and the 
cosmological constant$(\Lambda)$. The theory has also been found to 
overcome the shortcomings of the old inflationary theory
viz. the graceful exit problem \cite{a:p} and the longstanding problem of 
density perturbation \cite{r:p}.Further it has been observed \cite{rf:p} 
that 
the theory also preserves the generic features of Vilenkin \cite{v:p} and 
Hartle-Hawking \cite{jh:p} wave funtions. Finally it has also been shown 
that \cite{d:c} the theory admits wormhole solutions both for real and 
imaginary fields.

\par
Despite such a wide range of successful applications of the theory, the 
actual form of coupling of the matter field with gravity $(f(\phi))$ is 
not known a priory. One usually chooses it in an adhoc manner. 
Capozziello etal \cite{c:il} made an attempt to find the form of such 
coupling under the assumption that the Lagrangian of the induced theory 
of gravity admits Noether symmetry, which further restricts the 
form of the potential for the scalar field. In the Robertson-Walker model 
they have observed that\cite{c:il} $k\ne0$ imposes strong constraints in 
the form of 
coupling and the potential. The form of coupling thus obtained makes the 
Lagrangian degenerate and the form of the potential was found to be sixth 
order in the scalar field $\phi$, for which only trivial solution is 
admissible.

\par
When the Hessian determinant $W = 
|\frac{\partial^2{L}}{\partial{\dot{q_{i}}}\partial{\dot{q_{j}}}}|$ vanishes 
the Lagrangian 
becomes degenerate which imposes a constraint in the sense that the 
Legendre transformation does not exist and hence the Hamiltonian of the 
system can not be defined unless such constraints are analysed properly. 
In the domain of Lagrangian dynamics the constraint implies more number 
of degrees of freedom than the number of field equations, which means, 
one has to make certain assumptions to obtain exact solutions. However 
such degeneracy does not in any way lead to trivial solutions. This has 
been pointed out in a recent communication \cite{a:c}, where it has been 
shown that the existence of only trivial solutions is not due to 
the presence of degeneracy in the Lagrangian, rather due to the 
existence of Noether potential in the form 
$V(\phi)=\Lambda\phi^6$, which does not satisfy the field equations. 
This is a striking feature and perhaps not been encountered earlier. 
The reason for such contradiction, that the N$\ddot{o}$ther symmetry 
of the Lagrangian restricts the form of the potential in such a 
manner that it does not satisfy the field equations, is not known at 
present. However, it has been shown in the paper \cite{a:c} that for a 
choice of the coupling parameter in the form $f(\phi)= 
\epsilon\phi^2$, $\epsilon\ne -1/12$, which does not make the 
Lagrangian degenerate and for a quartic potential the Lagrangian 
admits certain dynamical symmetries, other than Noether 
symmetry, along with a conserved current. We emphasize on the fact 
that the symmetry thus obtained can not be explored by the 
standard technique of finding dynamical symmetries via 
Noether theorem. Rather, it is found simply from a 
combination of the field equations. This puts forward a vital 
question on the longstanding claim that all the dynamical 
symmetries of a physical system are Noether symmetries.

\par          
Motivated by the above mentioned result, our attempt is now to find the 
form of coupling $f(\phi)$ in models other than Robertson-Walker, under 
the same assumption that the Lagrangian admits Noether symmetry. 
In the present paper the Kantowski-Sachs metric has been taken under 
consideration.In this model, once again, we observe that Noether 
symmetry exists at the cost of imposing degeneracy in the Lagrangian. 
However, the potential this time turns out to be quartic in the scalar 
field $\phi$ which satisfies the field equations, in contrast to the 
Robertson-Walker model.The constraint imposed by the degeneracy has been 
analysed in the domain of Lagrangian dynamics, which has been found to 
yield an excess number of the degrees of freedom to the field equations. 
However, the most interesting aspect of the present work is that, instead 
of working with the whole lengthy process of finding Noether 
symmetry, it has been shown to obtain the same, just by inspecting the 
field equations. In this method the coupling parameter has 
been chosen in the form $f(\phi)=\epsilon\phi^2$ as in the 
Robertson-Walker model \cite{a:c}. It has been observed that the quartic 
form of the potential yields the Noether symmetry for 
$\epsilon=-1/12$. However, for any other arbitrary $\epsilon$ there 
exists yet another symmetry along with a conserved current, keeping the 
Lagrangian nondegenerate. This is surprising that such an inherent symmetry 
of the system can not be explored via Noether theorem. Hence we 
conclude that there exists dynamical symmetries of a system other than 
Noether symmetry. The symmetry thus obtained has got important 
consequences. Fakir, Unruh and Habib \cite{r:p} have shown that large 
negative $\epsilon$ may lead to well behaved self-consistant classical 
solutions which admit inflationary behaviour. Further in the framework of 
chaotic inflation, it can also produce density perturbations of amplitude 
consistant with the large scale behaviour, keeping the cosmological 
constant $\Lambda$ within the order $(10^-{2})$ of the ordinary GUT range. 
Since the symmetry that we have explored does not make any restriction on 
$\epsilon$ in general, so the results of Fakir,Unruh and Habib 
\cite{r:p} should also be realised, in principle, in this case too. 
Finally, that the large negative value of $\epsilon$ admits wormhole 
solution has already been confirmed in \cite{d:c}
\par
This paper is organised in the following manner. In section 2, the field 
equations are obtained from the action principle and the condition for 
which the Hessian determinant vanishes yielding a degenerate Lagrangian 
has been found. The form of coupling $f(\phi)$, the potential $V(\phi)$ 
and the conserved current are then obtained by studying the 
N$\ddot{o}$ther symmetry. It has been found that for the existence of 
such symmetry, the Lagrangian turns out to be degenerate. In section 3, 
the constraint imposed by the degenerate system is analysed in the domain 
of Lagrangian dynamics, whose outcome is a pair of field equations in 
first order for three degrees of freedom. This implies that one has to 
make one physically reasonable assumption to obtain nontrivial solutions. 
A solution has also been presented at the end of this section. In section 
4, it has been shown that the above mentioned symmetry could have been 
obtained quite easily just by inspecting the field equations. Further it 
has also been found that some other dynamical symmetry for the system 
still exists, that can not be obtained by applying the Noether 
theorem and that does not make the Lagrangian degenerate. Thus it is 
confirmed that not all the dynamical symmetries hidden in a Lagrangian 
could be obtained by the application of Noether's theorem. 
Concluding remarks are presented in section 5. 

\section{\bf{Noether Symmetry In Kantowski-Sachs Model}}

We start with the following action,
\be
A = \int~~d^4X \sqrt{-g}[f(\phi)R -\frac{1}{2}\phi,_{\mu}\phi'^\mu - 
V(\phi)] \ee
which for the Kantowski-sachs metric
\be
ds^2 = -dt^2 + a^2 dr^2+ b^2 (d\theta^2 + sin^2\theta d\phi^2)
\ee
reduces to
\be
A = 4\pi\int~~[-4f'ab\dot{b}\dot{\phi}-2f' 
b^2\dot{a}\dot{\phi}-4fb\dot{a}\dot{b}-2fa\dot{b}^2+2fa+\frac{1}{2}a 
b^2\dot{\phi}^2-a b^2 V(\phi) ]dt   + surface-term.
\ee
Field equations are
\be
2\frac{\ddot{b}}{b}+\frac{f'}{f}\ddot{\phi}+\frac{f''}{f}\dot{\phi}^2+2\frac{f'}{f}\frac{\dot{b}}{b}\dot{\phi}+\frac{\dot{b}^2}{b^2}+\frac{\dot{\phi}^2}{4f}+\frac{1}{b^2}-\frac{V(\phi)}{2f} = 0
\ee
\be
\frac{\ddot{a}}{a}+\frac{\ddot{b}}{b}+\frac{f'}{f}\ddot{\phi}+\frac{\dot{a}\dot{b}}{ab}+\frac{f'\dot{a}}{fa}\dot{\phi}+\frac{f'\dot{b}}{fb}\dot{\phi}+\frac{f''}{f}\dot{\phi}^2+\frac{\dot{\phi}^2}{4f}-\frac{V(\phi)}{2f}=0
\ee
\be
\frac{\ddot{a}}{a}+2\frac{\ddot{b}}{b}+2\frac{\dot{a}\dot{b}}{ab}+\frac{\dot{b}^2}{b^2}-\frac{\ddot{\phi}}{2f'}-(\frac{\dot{a}}{a}+2\frac{\dot{b}}{b})\frac{\dot{\phi}}{2f'}+\frac{1}{b^2}-\frac{V'(\phi)}{2f'}=0
\ee
\be
\frac{\dot{b}^2}{b^2}+\frac{f'\dot{a}}{fa}\dot{\phi}+2\frac{f'\dot{b}}{fb}\dot{\phi}-\frac{\dot{\phi}^2}{4f}+2\frac{\dot{a}\dot{b}}{ab}+\frac{1}{b^2}-\frac{V(\phi)}{2f}=0
\ee
where overdot and prime represent derivatives with respect to time and 
$\phi$ respectively. The Hessian 
determinant,$W=|\frac{\partial^2 
{L}}{\partial\dot{q_{i}}\partial\dot{q_{j}}}|$ turns out to be, \be
W = -16\pi fa b^4(3f'^2+f)
\ee
Hence, for $3f'^2+f=0$, whose exact solution is 
\be
f= -\frac{1}{12}(\phi-\phi_{o})^2
\ee
the Hessian determinant vanishes and the Lagrangian (3) becomes 
degenerate as in the Robertson-Walker case \cite{c:il} and \cite{a:c}.
\par
Let us now turn our attention to find the condition under which the 
Lagrangian(3) would admit Noether symmetry. In the Lagrangian 
under consideration the configuration space is $Q=(a,b,\phi)$,whose 
tangent space is $TQ=(a,b,\phi,\dot{a},\dot{b},\dot{\phi})$. Hence the 
infinitesimal generator of the Noether symmetry is 
\be
X = 
\alpha\frac{\partial}{\partial{a}}+\beta\frac{\partial}{\partial{b}}+\gamma\frac{\partial}{\partial{\phi}}+\dot{\alpha}\frac{\partial}{\partial\dot{a}}+\dot{\beta}\frac{\partial}{\partial\dot{b}}+\dot{\gamma}\frac{\partial}{\partial{\dot{\phi}}}
\ee 
The existence of Noether symmetry implies the existence of a 
vector field X such that the Lie derivative of the Lagrangian with 
respect to the vector field vanishes, ie.
\be
\pounds_{X}L = 0.
\ee
This yields an expression which is second degree in $a, b$ and $\phi$ and 
whose coefficients are functions of $a, b$ and $\phi$ only. Thus to satisfy 
equation [11], we obtain a set of following equations,
\be
2f\frac{\partial{\beta}}{\partial{a}}+f'b\frac{\partial{\gamma}}{\partial{a}}=0 
\ee
\be
\alpha+2b\frac{\partial{\alpha}}{\partial{b}}+2a\frac{\partial{\beta}}{\partial{b}}+a\frac{f'}{f}(\gamma+2b\frac{\partial{\gamma}}{\partial{b}})=o
\ee
\be
b\alpha+2a\beta+2ab\frac{\partial{\gamma}}{\partial{\phi}}-4f'(b\frac{\partial{\alpha}}{\partial{\phi}}+2a\frac{\partial{\beta}}{\partial{\phi}})=o
\ee
\be
\beta+b\frac{\partial{\alpha}}{\partial{a}}+a\frac{\partial{\beta}}{\partial{a}}+b\frac{\partial{\beta}}{\partial{b}}+b\frac{f'}{f}(\gamma+a\frac{\partial{\gamma}}{\partial{a}}+\frac{b}{2}\frac{\partial{\gamma}}{\partial{b}})=o
\ee
\be
f(b\frac{\partial{\alpha}}{\partial{\phi}}+a\frac{\partial{\beta}}
{\partial{\phi}})+f'(b\alpha+a\beta+\frac{b^2}{2}\frac{\partial{\alpha}}
{\partial{b}}+ab\frac{\partial{\beta}}{\partial{b}}+ab\frac{\partial{\gamma}}
{\partial{\phi}})+{f''}ab\gamma-\frac{ab^2}{4}\frac{\partial{\gamma}}{\partial
{b}}=o
\ee
\be
f\frac{\partial{\beta}}{\partial{\phi}}+f'(\beta+\frac{b}{2}\frac{\partial
{\alpha}}{\partial{a}}+a\frac{\partial{\beta}}{\partial{a}}+\frac{b}{2}
\frac{\partial{\gamma}}{\partial{\phi}})+\frac{{f''}b\gamma}{2}-\frac{ab}{4}
\frac{\partial{\gamma}}{\partial{a}}=o
\ee
\be
\alpha+\frac{f'}{f}a\gamma-\frac{ab^2}{2}[V(\frac{\alpha}{a}+2\frac{\beta}{b})+V'\gamma]=o
\ee
The above set of differential equations can essentially be solved by the 
method of separation of variables which finally yields a differential 
equation in f viz,
\be
3f'^2+f=o
\ee
whose solution is already given in equation (9). In addition 
$\alpha,\beta,\gamma$ and $V$ are also obtained in the process as
\be
\alpha=\frac{2l}{ab(\phi+\phi_{0})^3}, 
\beta=\frac{l}{a^2(\phi+\phi_{0})^3}, 
\gamma=-\frac{l}{a^2b(\phi+\phi_{0})^2}, V=\lambda(\phi+\phi_{0})^4 
\ee
where $l, \lambda$ and $\phi_{0}$ are constants of intregrations.
So the Lagrangian (3) admits Noether symmetry under the above 
condition that $f$ should have the form given by (9) while $V$ should be 
quartic in the scalar field $\phi$. However, it is to be noted that the 
form of $f$ given by (9) makes the Lagrangian degenerate. Thus a 
constraint has been imposed on the Lagrangian in order that it admits 
Noether symmetry.
\par
Now for Cartan one form
\be
\theta_{L}=\frac{\partial{L}}{\partial{\dot{a}}}da+\frac{\partial{L}}{\partial{\dot{b}}}db 
+\frac{\partial{L}}{\partial{\dot{\phi}}}d\phi 
\ee
the constant of motion $i_{X}\theta_{L}$ is obtained as,
\be
F=\frac{\frac{d}{dt}[b(\phi+\phi_{0})]}{a(\phi+\phi_{0})^2}=\frac{\frac{d}{dt}(b\phi)}{a\phi^2}   (for\phi_{0}=0)
\ee

\section{\bf{Analysing The Constraint And Presenting A Solution}}
The degeneracy in the Lagrangian imposed by the claim that it should 
have Noether symmetry, leads to constrained dynamics as mentioned 
in the introduction. This gives rise to underdetermined situation where the 
number of the true degrees of freedom exceeds the number of the field 
equations. To apprehend the situation, let us substitute $f, f', 
f''$ from equation (9) and $V, V'$ from equation (20) in the 
field equations (4-7) to obtain
\be
2\frac{\ddot{b}}{b}+2\frac{\ddot{\phi}}{\phi}+\frac{\dot{b}^2}{b^2}-\frac{\dot{\phi}^2}{\phi^2}+4\frac{\dot{b}\dot{\phi}}{b\phi}+\frac{1}{b^2}+6\lambda\phi^2=o
\ee
\be
\frac{\ddot{a}}{a}+\frac{\ddot{b}}{b}+2\frac{\ddot{\phi}}{\phi}+\frac{\dot{a}\dot{b}}{ab}+2\frac{\dot{a}\dot{\phi}}{a\phi}+2\frac{\dot{b}\dot{\phi}}{b\phi}-\frac{\dot{\phi}^2}{\phi^2}+6\lambda\phi^2=o
\ee
\be
\frac{\ddot{a}}{a}+2\frac{\ddot{b}}{b}+3\frac{\ddot{\phi}}{\phi}+2\frac{\dot{a}\dot{b}}{ab}+3\frac{\dot{a}\dot{\phi}}{a\phi}+6\frac{\dot{b}\dot{\phi}}{b\phi}+\frac{\dot{b}^2}{b^2}+\frac{1}{b^2}+12\lambda\phi^2=o
\ee
\be
\frac{\dot{b}^2}{b^2}+3\frac{\dot{\phi}^2}{\phi^2}+2\frac{\dot{a}\dot{\phi}}{a\phi}+2\frac{\dot{a}\dot{b}}{ab}+4\frac{\dot{b}\dot{\phi}}{b\phi}+\frac{1}{b^2}+6\lambda b^2=o
\ee
In addition we have yet another equation viz, equation (22), which is 
actually the constraint that has to be satisfied by the field equations 
(23-26).In order to see whether any new constraint arises from these 
field equations, we have to take time derivative of equation (22) and 
eliminate 
acceleration terms \cite{k:s} between the equation thus obtained and the 
field 
equations (22-26). Time derivative of equation (22) is (using the same 
equation in it)
\be
\frac{\ddot{b}}{b}+\frac{\ddot{\phi}}{\phi}=2\frac{\dot{\phi}^2}{\phi^2}+F\frac{\dot{a}\phi}{b}
\ee

Now eliminating acceleration terms between equations (23) and (27) one 
gets back the Hamiltonian constraint equation (26). Hence equation (13) 
is no longer an independent equation. In view of equations (22) and (26), 
one can obtain yet another constraint equation, viz,
\be
\frac{d}{dt}(a\phi)=-\frac{1+F^2{a^2}\phi^2}{2Fb}-\frac{3\lambda\phi^2{b}}{F}   
\ee
which can be used instead of equation (26). Differentiating equation (28) 
with respect to time and using the same equation once again in it, one 
obtains, 
\be
\frac{\ddot{a}}{a}+\frac{\ddot{\phi}}{\phi}=-2\frac{\dot{a}\dot{\phi}}{a\phi}+\frac{1+F^2{a^2{\phi^2}}}{2Fa{b^2}\phi}\dot{b}-\frac{3\lambda\phi}{Fa}\dot{b}-\frac{6\lambda b}{Fa}\dot{\phi}
+\frac{1+F^2{a^2}\phi^2}{2b^2}+3\lambda\phi^2
\ee
In view of equations (27) and (29) one can now easily observe that 
equations (24) and (25) are trivially satisfied. Hence at this stage we 
are left with a pair of equations viz, equations (22) and (28) with three 
degrees of freedom viz, $a, b$ and $\phi$, leading to an underdetermined 
situation. This is the outcome of a degenerate Lagrangian. In order to 
obtain solution, one is now free to impose `one' condition that would lead 
to physically acceptable solution. We are presenting here one such 
solution under the assumption, 
\be
a\phi=kb
\ee
wrere $k$ is a constant and let it be positive definite $(k>0)$. In view 
of equation (30) equation (22) can immediately be integrated to yield, 
\be
b\phi=n\exp{(Fkt)}
\ee
where $n$ is a constant of integration and considered to be positive 
definite $(n>0)$ too. Further for $\lambda = 0$, equation (28) can also be 
integrated in view of equation (30). the result is 
\be
b=\frac{1}{Fk}[m\exp{(-Fkt)}-1]^{\frac{1}{2}} 
\ee
where the overall negative sign has been chosen to reveal physically 
acceptable solution and m is yet another constant of integration which is 
considered to be greater than one $(m>1)$ for the same reason. Hence 
$\phi$ and $a$ can also be obtained in view of equations (30), (31) and 
(32) as,
\be
\phi=-nFk\frac{\exp{(Fkt)}}{[m\exp{(-Fkt)}-1]^\frac{1}{2}}  ,  
a=\frac{\exp{(-Fkt)}}{nkF^2}[m\exp{(-Fkt)}-1] \ee
Now if one chooses $F=-c^2$, then the solutions (32) and (33) take the 
following form,
\be
a=\frac{\exp{(c^2{kt})}}{nkc^4}[m\exp{(c^2{kt})-1}], 
b=\frac{1}{kc^2}[m\exp(c^2{kt})-1]^\frac{1}{2},  
\ee
\be
ab^2=\frac{\exp{(c^2{kt})}}{nk^3{c^8}}[m\exp(c^2{kt})-1]^2, 
\phi=nkc^2\frac{\exp(-c^2{kt})}{[m\exp(c^2{kt})-1]^\frac{1}{2}} \ee
The above solution reveals that the universe admits inflation starting 
from a finite proper volume, under the choice of the constants already 
made viz, $k>0, n>0$ and $m>1$. the scalar field at the initial epoch is 
finite and it falls off exponentially as the universe expands.The 
solution is singularity free although there is no question of graceful 
exit from inflation. the big-bang singularity is pushed back to the infinite 
past.  

\section{\bf{Some Other Symmetries In Kantowski-Sachs Model}}

In this section we shall first show that the whole laborious job that has 
been carried out in the preeciding section, to find the set of equations 
(12) to (18)  and to solve them by the method of the separation of variables 
to obtain conditions under which the Lagrangian (3) admits 
Noether symmetry, is not at all required. Rather we can construct 
a pair of equations from the set of field equations (4) to (7), one of 
which can immediately extract the conditions for which the Lagrangian 
would admit dynamical symmetry and the other can find the corresponding 
conserved current. We shall further show that one of the dynamical 
symmetries obtained in the process is of Noether class and there 
exists dynamical symmetries of some other type that we could not obtain by 
applying Noether's theorem in the preceeding section.  
\par
The first one of this pair is the continuity equation. This equation is 
obtained by eliminating $\ddot{a}$ and $\ddot{b}$ from the field equations 
(4) to (6) and then comparing it with the Hamiltonian constraint equation 
(7). The equation thus formed is,
\be
2(3f'^2+f)(\ddot{\phi}+\frac{\dot{a}}{a}\dot{\phi}+2\frac{\dot{b}}{b}\dot{\phi})+f'(6f''+1)\dot{\phi}^2+2(fV'-2Vf')=0.
\ee
All dynamical symmetries are hidden in this equation. To find 
Noether symmetry one has to choose $f$ and $V$ in such a way 
that equation (36) is satisfied identically. The choice is quite trivial 
viz, the coeffients of the derivatives of $\phi, a$ and $b$ should vanish 
separately. This implies $3f'^2 + f =0$ ie, $f = 
-\frac{1}{12}\phi^2$ and as such $6f'' + 1 = 0$ too. From the last 
term of equation (36) one can see that $V$ is proportional to $f^2$ and 
hence is in the form $V = \lambda \phi^4$. These results are already 
obtained in equations (19) and (20) of the preceeding section. To obtain 
the conserved current we construct yet another equation and that is done 
simply by eliminating terms in the field equations which are free from 
time derivatives viz, $\frac{1}{b^2}, V(\phi)$ and $V'(\phi)$ in the 
present 
context. This is done by taking the difference of equations (1) and (4), 
which yields,
\be
2\frac{\ddot{b}}{b}+\frac{f'}{f}\ddot{\phi}+(\frac{2f''+1}{2f})\dot{\phi}^2-\frac{f'\dot{a}}{fa}\dot{\phi}-2\frac{\dot{a}\dot{b}}{ab}=0
\ee
This equation in view of the solution of $f$ obtained from equation (36), 
reads
\be
\frac{\ddot{b}}{b}+\frac{\ddot{\phi}}{\phi}-2\frac{\dot{\phi}^2}{\phi^2}-\frac{\dot{a}\dot{\phi}}{a\phi}-\frac{\dot{a}\dot{b}}{ab}=0
\ee
whose first integral yields the conserved current obtained in equation 
(22). Thus, we have shown that the Noether symmetry can even be 
obtained in view of the continuity equation (36) and the corresponding 
conserved current from equation (37), without invoking equations (11) and 
(21). 
\par
Let us now proceed to find some other type of dynamical symmetry that we 
did not find in the preceeding section, for which we choose $f$ in the form 
\be
f=\epsilon \phi^2
\ee
Further we choose the potential in the form
\be
V=\lambda \phi^4
\ee
then equation (36) can be written as 
\be
(12\epsilon+1)(\ddot{\phi}+\frac{\dot{a}}{a}\dot{\phi}+2\frac{\dot{b}}{b}\dot{\phi}+\frac{\dot{\phi}^2}{\phi})=0
\ee
For $\epsilon=-1/12$, we regain Noether symmetry. However, for 
any arbitrary $\epsilon$ other than zero or $(-1/12)$, equation (41) leads 
to 
\be
\frac{\ddot{\phi}}{\phi}+\frac{\dot{a}\dot{\phi}}{a\phi}+2\frac{\dot{b}\dot{\phi}}{b\phi}+\frac{\dot{\phi}^2}{\phi^2}=0
\ee
whose first integral is
\be
ab^2 \phi\dot{\phi}=constant
\ee
Thus we obtain yet another dynamical symmetry of the system for arbitrary 
value of $\epsilon$, for which the conserved current is given by equation 
(43).It is to be noted that the existence of this dynamical symmetry does 
not make the Lagrangian degenerate, since the Hessian determinant given 
by equation (8) does not turn out to be zero. Since this symmetry has not 
been obtained by the application of Noether's theorem, therefore 
we conclude that not all the dynamical symmetries of a Lagrangian are of 
Noether class. The dynamical symmetry thus obtained  here is of 
the same form that we have already seen in connection with the 
Robertson-Walker metric \cite{a:c}. This type of symmetry is of much 
interest. Since the symmetry exists for arbitrary $\epsilon$, it exists 
for conformally coupled scalar field $\epsilon = 1/6$ as well. Further as 
already mentioned in the introduction that large negative value of 
$\epsilon$ might give rise to well behaved self consistent classical 
solution admitting inflation \cite{r:p} on one hand and can produce density 
perturbations of amplitude consistent with the large scale behavior 
keeping the cosmological constant $\lambda$ within the ordinary GUT 
range, on the other, therefore the existence of dynamical symmetry 
corresponding to arbitrary value of $\epsilon$  is definitely of much 
importance.
\section{\bf{Concluding Remarks}}
 In a recent communication we have come across an important and wonderful 
result, while reviewing the works of Capozziello et al \cite{c:il} in 
connection with the Noether symmetry in the Robertson-Walker 
metric. The result is that, even if there exists certain forms of 
$f(\phi)$ and $V(\phi)$ and hence a vector field $X$ such that 
$\pounds_{X}L=0$: the form of $V(\phi)$ might not satisfy the field 
equations. We have not come across such a result earlier and as such 
do not know the reason as yet. However, we have observed that only one, 
viz, the continuity equation suffice to check wheather the field 
equations are satisfied or not. In the context of the induced theory 
of gravity, this equation turned out to be a very important one to 
explore all sorts of existing dynamical symmetries of a Lagrangian.        
\par
In the present paper, we have shown that the Noether symmetry of 
the induced theory of gravity in the Kantowski-Sachs model can be traced 
from the continuity equation, while instead of using the Cartan's one 
form, the conserved current can simply be found from equation (37) or 
(38), which is obtained from yet another combination of the field 
equations. Though the Noether symmetry makes the Lagrangian 
degenerate and hence introduces a constraint which reduces the number of 
independent field equations to the number of true degrees of freedom by 
one, causing underdeterminancy, yet the field equations are found to admit 
inflationary solution under a suitable assumption. This result definitely 
proves that the conclusion made by Capozziello et al \cite{c:il}, viz, 
degeneracy leads to trivial solutions, is wrong.
\par 
It has been further observed that all sorts of dynamical symmetries of a 
Lagrangian can be explored from the continuity equation only, at least in 
induced theory of gravity. This equation further reveals dynamical 
symmetries other than the Noether symmetry, for a Lagrangian.  
This confirms that not all the dynamical symmetries of a system belong to 
the Noether class.

\end{document}